\documentclass[aps,prb,twocolumn,groupedaddress]{revtex4}
\usepackage{graphicx}
\usepackage{color}
\usepackage{epstopdf}
\usepackage{epsfig}
\begin{document}

\title{Quasiparticle diffusion based heating in superconductor tunneling micro-coolers}

\author{Sukumar Rajauria, Herv\'e Courtois and Bernard Pannetier}
\affiliation{Institut N\'eel, CNRS and University Joseph Fourier, 25 Avenue des Martyrs, 38042 Grenoble Cedex 9, France}
\date{\today}

\begin{abstract}
In a hybrid Superconductor - Insulator - Normal metal tunnel junction biased just below the gap, the extraction of hot electrons out of the normal metal results in electronic cooling effect. The quasiparticles injected in the superconductor accumulate near the tunnel interface, thus increasing the effective superconductor temperature. We propose a simple model for the diffusion of excess quasiparticles in a superconducting strip with an additional trap junction. This diffusion model has a complete analytic solution, which depends on experimentally accessible parameters. We find that the accumulated quasiparticles near the junction reduce the efficiency of the device. This study is also relevant to more general situations making use of superconducting tunnel junctions, as low temperature detectors.
\end{abstract}
\maketitle

In a Superconductor (S) - Insulator (I) - Normal metal (N) junction biased below the superconducting energy gap $\Delta$, the quasiparticle tunnel current transfers heat from the normal metal to the superconductor. This results in a significant cooling of the normal metal electronic population.\cite{review} In most cases, the cooling is less efficient than expected, in particular near the optimal bias just below the gap. It is often argued that a parasitic power could explain this behaviour. A more intrinsic explanation is local overheating of the superconducting electrode, as excess quasiparticles with an energy close to the gap accumulate near the tunnel contact and can tunnel back into the normal metal. The use of quasiparticle traps on the superconducting electrode has been found to improve the cooling performance.\cite{traps} A quantitative description of the overheating effects calls for a precise analysis of the decay mechanisms of the excess quasiparticle population in the superconducting electrode. In this framework, a detailed theory of non-equilibrium phenomena in a micro-cooler with normal metal traps on the superconducting electrodes was recently developed.\cite{heikkila}

This paper is an attempt to describe the low temperature diffusion of non-equilibrium quasiparticles in a superconducting strip coupled to a normal metal trap, in the presence of recombination and pair-breaking processes. Our model is an extension of the Rothwarf-Taylor equations \cite{RT} for the relaxation of non-equilibrium quasiparticles. A normal metal trap junction provides an additional channel for near-gap quasiparticle evacuation. Under realistic assumptions, the non-linear diffusion equation can be solved analytically. The spatial profile of the quasiparticle population in the S strip is obtained. Near the tunnel junction, the quasiparticle density exceeds the thermal equilibrium density, even at low injection current. The heat transferred back to the normal electrode is found to reduce the effective cooling power of the tunnel junction. We conclude by comparing our predictions to experimental results. 

\begin{figure}[t]
\centering
\includegraphics[width=8.5cm]{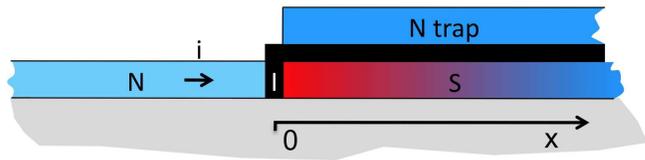}
\caption{Schematic cut view of the system considered. The excess quasiparticles injected from the N-metal in the S electrode diffuse out of the junction region and relax by two processes: recombination and escape in the normal metal trap. The trap layer is supposed to overlap completely the superconducting strip and both have infinite length. Both the substrate and the trap layer are at the bath temperature is $T_{0}$.}
\label{model}
\end{figure}

The system we consider is the superconducting section of a low temperature N-I-S cooler (Fig. \ref{model}). It is modeled by a semi-infinite line in contact with a dielectric substrate. The superconductor is covered by a second normal metal $N_{trap}$ separated by a tunnel barrier. A quasiparticle current is injected through a N-I-S junction at the origin of the x-axis. We assume that the layers thicknesses are small, so that we consider diffusion only along the x-axis. Practical micro-coolers are usually of a S-I-N-I-S geometry, i.e. they are symmetrically made of a double N-I-S junction. The geometry under study here thus corresponds to one half of a micro-cooling device. 
 
At zero current, the system is at equilibrium at the bath temperature  $T_{0}$, assumed to be much smaller than the superconductor critical temperature. Upon increasing the current in the N-I-S cooler, the electron temperature $T_{N}$ of the N-metal decreases while the quasiparticle population increases on the superconductor side. If the cooler junction is biased below the gap value, only the high energy tail of the normal metal electron population can tunnel. Accordingly, the distribution of injected quasiparticles is sharply peaked at the gap energy $\Delta$. Its width of the order $k_{B} T_{N} \ll \Delta$ ($k_{B}$ is the Boltzman constant) will be ignored in the following.

The process of recombination in a non-equilibrium superconductor was discussed by Rothwarth and Taylor \cite{RT} for thin film geometries. The decay of the quasiparticle density involves quasiparticle recombination retarded by phonon retrapping. In this process, two quasiparticles initially recombine to form a Cooper pair, resulting in the creation of a phonon of energy $2\Delta$. This phonon can be subsequently reabsorbed by a Cooper pair, resulting in two new quasiparticles. The $2\Delta$ energy leaves the superconductor when either the phonons or the quasiparticles escape to the bath. In our model, we introduce two additional contributions to the Rothwarth-Taylor equations: the diffusion of quasiparticles in the superconducting electrode and their absorption by the trap junction. In the superconducting strip, the system of equations for the density of quasiparticles per unit volume $N_{qp}$ and the density of 2$\Delta$ phonons $N_{2\Delta}$ then reads:
\begin{eqnarray}
\{ \frac{d}{dt}-D_{qp}\frac{d^2}{dx^2} \} N_{qp}=-RN_{qp}^2+\frac{2N_{2\Delta}}{\tau_{B}}-\frac{N_{qp}-N_{qp0}}{\tau_{0}}\\
\frac{dN_{2\Delta}}{dt}=\frac{1}{2}RN_{qp}^2 -\frac{N_{2\Delta}}{\tau_{B}}-\frac{N_{2\Delta}-N_{2\Delta0}}{\tau_{\gamma}}.
\label{RTequations}
\end{eqnarray}
Here $D_{qp}$ is the quasiparticle diffusion constant. It differs from the free electron diffusion constant because of the reduced group velocity of quasiparticles as compared to the Fermi velocity.\cite{Nara} There is no source term in Eq. 1 and 2 as the source will be introduced as a boundary condition for the density gradient at coordinate $x=0$.

In Eq. 1, the first term in the right hand side is the recombination term. Its quadratic dependence indicates that the recombination is a two-particle process.  The prefactor $R$ is a constant depending on the electron-phonon interaction. The related recombination time $\tau_{R}^{-1}=2 R N_{qp0}$ is strongly dependent on the temperature. At low temperature, it diverges as $\exp{(\Delta/T_{0})}$.\cite{Prober} The second term describes the quasiparticle creation by pair-breaking processes induced by a $2 \Delta$ phonon (two quasiparticles per phonon). The related phonon pair-breaking time $\tau_{B}$ is weakly dependent on temperature. $\tau_{R}$ and $\tau_{B}$ are intrinsic characteristic times of the superconducting material ruled by the electron-phonon interaction.\cite{kaplan} 

Eq. 2 has similar terms for the reverse processes but without gradient term because the $2\Delta$ phonon absorption time $\tau_{B}$ is very short.\cite{kaplan} The low energy (compared to $2\Delta$) phonons are ignored since their interaction with the superconducting condensate is negligible in a BCS superconductor at low temperature.\cite {pannetier77} The trap film is assumed to be at the bath temperature. The escape terms on the right hand side of each equation describe respectively the quasiparticle escape by tunneling to the trap film at a rate $\tau_{0}^{-1}$ and the phonon escape to the substrate at a rate $\tau_{\gamma}^{-1}$. The characteristic time $\tau_{0}=R_{trap}e^{2}N(E_{F})d_{S}$ is proportional to the thickness $d_{S}$ of the S-film, the electron density of states at the Fermi level $N(E_{F}$) and the specific resistance of the trap tunnel junction $R_{trap}$. The escape time $\tau_{\gamma}$ for 2$\Delta$ phonons to the substrate depends on the acoustic matching at the interface to the substrate and on the superconducting film thickness.

With no injection current,  Eq. 2 reduces to a simple relation between the equilibrium populations (labelled with a "0" index): $RN_{qp0}^{2}=2N_{2\Delta 0}/\tau_{B}$. Following the BCS theory, $N_{qp0}$ decays exponentially at low temperature $T_{0}$ as:
\begin{equation}
N_{qp0} = N(E_{F})\Delta~\sqrt{\frac{\pi~k_BT_{0}}{2\Delta}}\exp[-\frac{\Delta}{k_BT_{0}}].
\label{qpdensity}
\end{equation}

In the following, we will use the dimensionless relative density $z(x)$ defined by:
\begin{equation}
N_{qp}(x)=N_{qp0}[1+z(x)]
\label{deltaN}
\end{equation}
Under steady state injection, Eq. 2 can be rewritten to express the density $N_{2 \Delta}$ as a function of $z(x)$. Eq. 1 then gives the following second order differential equation for $z(x)$:
\begin{equation}
D_{qp}\frac{d^{2}z}{dx^{2}}=\frac{z}{\tau_{0}}+\frac{z+z^{2}/2}{\tau_{eff}}.
\label{differential}
\end{equation}
where $\tau_{eff}=\tau_{R}(1+\tau_{\gamma}/\tau_{B})$ is an effective quasiparticle recombination time.\cite{RT} In particular, the situation $\tau_{\gamma} > \tau_{B}$ corresponds to $2\Delta$ phonons being re-absorbed before leaving the superconductor, which enhances the quasiparticle density.

The boundary conditions at the two wire ends are:
\begin{eqnarray}
z&=&0 \hspace{2cm}  x\rightarrow\infty \nonumber\\
-D_{qp}N_{qp0}\frac{dz}{dx}&=&\frac{i}{eA} \hspace{2cm}  x=0 \nonumber
\end{eqnarray}
The first one states that the excess quasiparticle density must decay to zero at the right end of S. The second one is the continuity condition for the quasiparticle flux at the injection point $x=0$: the gradient of the quasiparticle density is equal to the rate of quasiparticle injection per unit area. Here $A$ is the junction area and $i$ is the bias current. The injected electrons create quasiparticle excitations with energy about $\Delta$ per electron, and no charge, the electrical continuity being insured by the supercurrent in S. Charge imbalance effects can be ignored here since the N-I-S junction is biased near the gap edge.\cite{Tinkham72}

The solution of Eq. \ref{differential} gives the analytic expression of $z$ as a function of the $x$-coordinate:
\begin{equation}
\label{profi}
z(x)=\frac{6 \alpha}{\cosh[(x+a)/\lambda]-1 }
\end{equation}
It includes an integration constant $a$ which depends on the injection current $i$. It is obtained by substituting $z(x=0)$ in the following equation:
\begin{equation}
\label{bc}
z(0)\sqrt{1+\frac{z(0)}{3 \alpha}}=\frac{|i| \lambda}{eAD_{qp} N_{qp0}}=I_{inj}
\end{equation}
Here we have introduced the parameter $\alpha=1+\tau_{eff}/\tau_{0}$, which is the enhancement ratio of the quasiparticle decay rate due to the presence of the trap junction. The quasiparticle diffusion length $\lambda=\sqrt{D_{qp}\tau_{eff}/\alpha}$ combines the two mechanisms of absorption/recombination with phonons and trapping in the normal metal trap. When trapping is dominant, the decay length tends to $\sqrt{D_{qp}\tau_{0}}$. We also defined the dimensionless current $I_{inj}$ defined as the density of quasiparticles injected in the volume $\lambda A$ during the diffusion time $\lambda^2 / D_{qp}$.

\begin{figure}[t]
\centering
\includegraphics[width=8.5 cm]{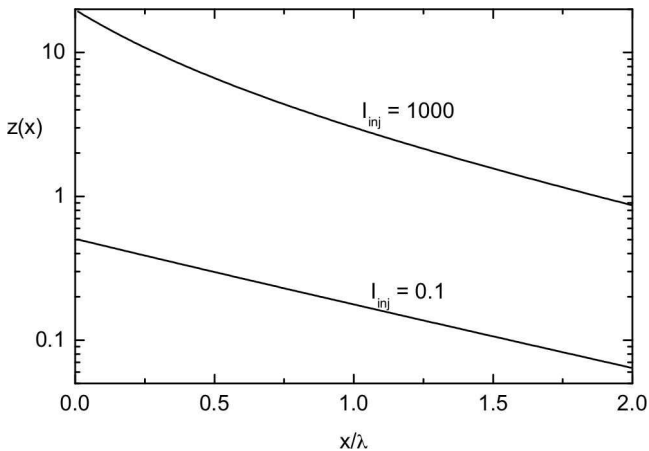}
\caption{The dimensionless relative density of quasiparticles $z(x)$ in logarithmic scale for two injection levels. The dimensionless current $I_{inj}$ is equal to respectively $0.1$ (lower trace) and $1000$ (upper trace). The horizontal axis is the $x$ coordinate is in units of the diffusion length $\lambda$. The $\alpha$ parameter is taken equal to $1$.}
\label{profile}
\end{figure}

At low injection current, the above equations can be linearized. The excess in quasiparticle density is then directly proportional to the injection current: $z(0) \approx I_{inj}$. The density $N_{qp}$ decays exponentially with the distance from the interface, which gives a linear behavior on a log scale, see Fig. \ref{profile} lower trace. The decay characteristic decay length is the diffusion length $\lambda$.

At high current, i.e. when the injected quasiparticle density is larger than the thermal one, Eq. \ref{bc} gives $z(0) \approx (3 \alpha I_{inj})^{2/3}$. This non-linearity originates from the recombination term $N_{qp}^{2}$ term in Eq. 2. The constant $a$ is then small compared to 1 and the decay turns to a power law $1/(x+a)^2$ (see Fig. \ref{profile} upper trace). The crossover current between the two current regimes, $i\approx \alpha D_{qp} N_{qp0}eA/\lambda$, decreases with the thermal quasiparticle density. Therefore, the non-linearity persists even at low current when the temperature is low. 

Let us now discuss how the extra quasiparticles affect the operation of a N-I-S cooler. The presence of excess quasiparticles in the superconducting electrode has no direct signature on the current voltage characteristic of the tunnel junction. Essentially, the tunnel current is independent of the quasiparticle population in the superconductor and can be written as an integral over energy involving only the normal metal distribution function $f_{N}$:
\begin{equation}
i=\int_{0}^{\infty}\frac{n(E)dE}{er_{N}}[f_{N}(E-eV)-f_{N}(E+eV)].
\end{equation}
Here $n(E)$ is the BCS density of states ratio and $r_{N}$ is the normal state resistance of the cooler junction. Only under strong injection can a change in the I-V curve associated with the gap suppression be observed. In contrast, the heat current through the junction is an explicit function of the distribution function $f_{S}$ in the superconducting side:
\begin{equation}
J_{Q}=\int_{0}^{\infty}\frac{n(E)EdE}{e^{2}r_{N}}[2f_{S}(E)-f_{N}(E-eV)-f_{N}(E+eV)].
\label{JQ}
\end{equation}
where both forward tunneling and back-tunneling processes contribute.

In the usual analysis of the N-I-S junctions cooling power, the superconductor heating is ignored and therefore the distribution function $f_{S}$ is taken to be equal to the Fermi function at the bath temperature $T_{0}$. In order to introduce the contribution of excess quasiparticles, one can include the actual non-equilibrium population in $f_{S}$. As this population is peaked at the gap energy,  it is sufficient to consider an additional heat current equal to $\Delta$ times the tunneling rate, so that:
\begin{equation}
\label{backtunneling}
\delta P_{bt}=\frac{2\Delta }{e^2 r_{N}}\frac{N_{qp}(0)-N_{qp0}}{N(E_{F})}=\frac{2\Delta }{e^2 r_{N}}\frac{N_{qp0}}{N(E_{F})}z(0).
\end{equation}
$\delta P_{bt}$ is the electronic thermal current due to the back-tunneling of the excess quasiparticles at $x=0$. It is proportional to the excess quasiparticle density $N_{qp}(0)-N_{qp0}$ due to injection through the N-I-S junction. Besides the quasiparticle back-tunneling, there is also a phonon contribution due to the reabsorption of  $2\Delta$ phonons in the normal electrode.\cite{Jochum} The excess $2\Delta$ phonon density at position $x=0$ can be obtained as a function of $N_{qp}(0)$ from Eq. 2. It leads to a heat current $\sim (N_{qp}(0)-N_{qp0})^{2}$. It is a second order process and will be ignored in the following.

In a previous work,\cite{sukumar07} we studied the electronic cooling of the central N-island in a S-I-N-I-S junction. The normal island electronic temperature was extracted from the experimental current-voltage characteristics of the cooler junction. As expected, the experimental data show a minimum in electronic temperature at a bias just below the superconductor gap, when the heat current (Eq. \ref{JQ}) is maximum. The cooling power was analyzed within a thermal model combining electron-phonon interactions in the normal metal and phonon transfer through the interface. At the optimum bias $V\approx \Delta/e$, the electronic temperature was found to be systematically higher than predicted. For example, at bath temperature of 292 mK, the lowest electronic temperature was about 97 mK whereas the prediction is 58 mK (see Fig. 4 of Ref. \onlinecite{sukumar07}). We checked that the residual contributions due to imperfect filtering, leakage resistance or Andreev heating\cite{sukumar08} do not contribute significantly to the heating at the gap bias.

Let us now examine how the heat losses due to the quasiparticle back-tunneling $\delta P_{bt}$ could explain the observed discrepancy. We estimate $\delta P_{bt}$ from Eq. \ref{backtunneling} where the quasiparticle density has been substituted by its linear approximation in Eq. \ref{bc}. Within this approximation, the back-tunneling power is directly proportionnal to the tunnel current absolute value $|i|$:
\begin{equation}
\delta P_{bt}=f|i|\approx \frac{2\Delta \lambda}{e^{2} r_{N}N(E_{F})D_{qp}eA}|i|.
\end{equation}
The physical parameters of the device are a junction area $A$ = 1.5 $\times$ 0.3 $\mu$m$^{2}$, Cu and Al thickness $d_{N}$ = 50 nm and $d_{S}$ = 40 nm respectively, and the normal state resistance $r_{N}$ = 1.25 k$\Omega$. The diffusion coefficient of the Al film was measured at 4.2 K to be 30 cm$^{2}$/s. The devices were designed so that every Al electrode is overlapped by a Cu strip, which plays the role of a normal metal trap. As a first analysis, we neglect its separation with the injection junction (0.3 $\mu$m) compared to the quasiparticle diffusion length ($\approx$ 10 $\mu$m), so that we apply our simple analytical approach. The Al oxide barrier is identical in the trap and the cooler tunnel junctions, thus we can assume that their specific tunnel resistance is the same: $R_{trap}=r_{N} A$. Using $N(E_{F})$ = 3.44 $\times$ 10$^{10}$ eV$^{-1}\mu$m$^{-3}$,\cite{Kittel}  and assuming that the Cu trap overlaps the superconductor over about half of its width, we find $\tau_{0}$ = 0.4 $\mu$s. This is much smaller than the effective recombination time $\tau_{eff}$ for aluminum.\cite{Prober} It thus determines the effective diffusion length $\lambda \approx \sqrt{D_{qp} \tau_{0}}$ over which the quasiparticle energy is evacuated to the bath. If the free electron diffusion coefficient is used for $D_{qp}$, we estimate the parameter $f$ to be about 1 pW$/\mu$A.

\begin{figure}[t]
\centering
\includegraphics[width=8.5 cm]{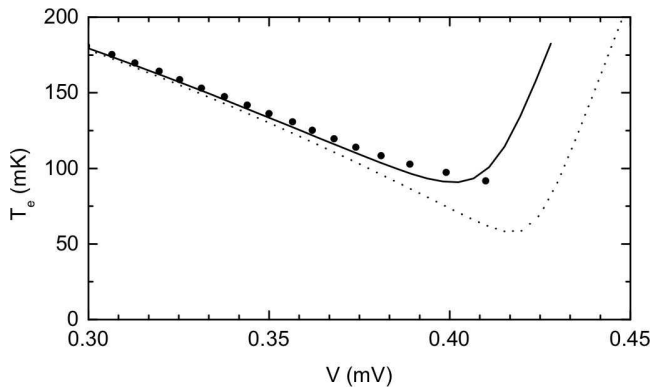}
\caption{Electronic temperature as a function of voltage bias in a S-I-N-I-S micro-cooler at a bath temperature of 292 mK. The data (circular dots) are fitted by the model described in Ref. \onlinecite{sukumar07} (dotted line). The continuous line is obtained by adding the superconductor heating effect with the parameter $f =$ 14 $pW/\mu A$.}
\label{fit}
\end{figure}

We have added the contribution of $\delta P_{bt}=f |i|$ to the thermal balance equation discussed in Ref. \onlinecite{sukumar07}, using $f$ as a fitting parameter and $i$ equal to the experimental current. We found that for the bath temperatures 292 (see Fig. \ref{fit}) and 489 mK, the agreement with the experiment could be restored over the whole subgap bias voltage range by choosing $f =$ 14 pW/$\mu$A. This is significantly larger than the calculated value. Still, we know that the latter is underevaluated for several reasons. First, we have overestimated $D_{qp}$ by taking the normal state value, which ignores the vanishing group velocity of quasiparticles near the gap energy. The reduction factor varies as $kT_N /\Delta$ at very low temperature. Since $P_{bt}$ is proportional to $\lambda/D_{qp}$, it must be multiplied by the inverse square root of this reduction factor, which amounts to 5 at $0.1K$. Second, we have ignored the uncovered part of the superconducting electrode nearby the cooler junction. Eventually, alternative explanations of the efficiency mismatch at the gap edge remain fully possible. 

In summary, we developed a simple model for the diffusion of excess quasiparticles in a S-strip, including the effect of an external trap. Our assumption of a superconducting line with an infinite length leads to simple and exact expressions. In more complicated designs used in experimental realizations, a numerical solution would be necessary. Our model can describe unexplained values of the minimum temperature observed in superconducting micro-coolers. It should also be useful to test accurately the influence of material and geometrical parameters and hence improve the performance of micro-coolers and superconducting tunnel junctions based low temperature detectors. 

The authors acknowledge useful discussions with F. W. J. Hekking, L. Pascal, L. Swenson and T. T. Heikkil\"{a}. This work was supported by the R\'egion Rh\^one-Alpes and the NanoSciERA project "Nanofridge".

\end{document}